\documentclass[12pt]{article}   
\usepackage{psfig,times,fancyheadings}  

\begin{document}
\thispagestyle{empty}

 \lhead[\fancyplain{}{\sl }]{\fancyplain{}{\sl }}
 \rhead[\fancyplain{}{\sl }]{\fancyplain{}{\sl }}

 \renewcommand{\topfraction}{.99}      
 \renewcommand{\bottomfraction}{.99} 
 \renewcommand{\textfraction}{.0}


\newcommand{\nc}{\newcommand}

\nc{\qI}[1]{\section{{#1}}}
\nc{\qA}[1]{\subsection{{#1}}}
\nc{\qun}[1]{\subsubsection{{#1}}}
\nc{\qa}[1]{\paragraph{{#1}}}

\def\qbu{\hfill \par \hskip 6mm $ \bullet $ \hskip 2mm}
\def\qee#1{\hfill \par \hskip 6mm #1 \hskip 2 mm}

\nc{\qfoot}[1]{\footnote{{#1}}}
\def\qL{\hfill \break}
\def\qpar{\vskip 2mm plus 0.2mm minus 0.2mm}
\def\qtvi{\vrule height 2pt depth 5pt width 0pt}
\def\qth{\vrule height 12pt depth 0pt width 0pt}
\def\qtb{\vrule height 0pt depth 5pt width 0pt}
\def\tvi{\vrule height 12pt depth 5pt width 0pt}

\def\qparr{ \vskip 1.0mm plus 0.2mm minus 0.2mm \hangindent=10mm
\hangafter=1}

\def\qdec#1{\par {\leftskip=2cm {#1} \par}}

\def\qdpt{\partial_t}
\def\qdpx{\partial_x}
\def\qddpt{\partial^{2}_{t^2}}
\def\qddpx{\partial^{2}_{x^2}}
\def\qn#1{\eqno \hbox{(#1)}}
\def\qds{\displaystyle}
\def\qw{\widetilde}
\def\qmax{\mathop{\rm Max}}   
\def\qmin{\mathop{\rm Min}}   


\def\qci#1{\parindent=0mm \par \small \parshape=1 1cm 15cm  #1 \par
               \normalsize}

\null

\centerline{\bf \Large The conundrum of stock versus bond prices}
\vskip 0.5cm
 \centerline{\bf \Large }

\vskip 1cm
\centerline{\bf Sergei Maslov $ ^1 $ and Bertrand M. Roehner $ ^2 $ }
\vskip 4mm

\vskip 2cm

{\bf Abstract}\quad In a general way, stock and bond prices do not
display any significant correlation. Yet, if we concentrate our attention
on specific episodes marked by a crash followed by a rebound,
then we observe that stock prices have a strong connection with
interest rates on the one hand, and with bond yield spreads on the 
other hand. That second relationship is particularly stable in the
course of time having been observed for over 140 years. 
Throughout the paper we use a quasi-experimental approach. By observing
how markets respond to well-defined exogenous shocks (such as
the shock of September 11, 2001) we are able to determine how investors
organize their ``flight to safety'': 
which safe haven they select, how long their collective panic lasts,
and so on. 
As rebounds come to an end
the correlation of stock and bond prices fades away, a clear sign that
the collective behavior of investors loses some of its coherence; this
observation can be used as an objective
criterion for assessing the end of a market rebound.
Based on the behavior of investors, we 
introduce a distinction between ``genuine stock market rallies'', as opposed
to spurious rallies such as those brought about by the buyback 
programs implemented by large companies. The paper ends with a discussion
of testable predictions.

\vskip 1cm

\centerline{25 October 2003}

\vskip 8mm
\vskip 8mm

\vskip 2cm
1: Department of Physics,  Brookhaven National Laboratory, Upton, 
New York 11973; maslov@bnl.gov
\vskip 4mm
2: Permanent affiliation: Institute for Theoretical and High Energy Physics,
University of Paris. 
\qL
\phantom{1: }Postal address to which correspondence should be sent:\qL
\phantom{1: }B. Roehner, LPTHE, University Paris 7, 2 place Jussieu, 
75005 Paris, France.
\qL
\phantom{1: }E-mail: roehner@lpthe.jussieu.fr
\qL
\phantom{1: }FAX: 33 1 44 27 79 90
\qL
\phantom{1: }{\it Most of this study was done during a stay this 
author made
as a visiting researcher at Brookhaven Lab in April-May 2003.}

\vfill \eject

\qI{Introduction}

Stocks and bonds constitute the two main securities traded on stock
exchanges. It is true that over the last two decades other products
such as futures and options have acquired an ever increased importance;
in a sense, however, one can consider that their main purpose is to
provide hedging tools for those ``primary products'' such as 
stocks, bonds, commodities, exchange rates and so on. It is therefore
a natural question, to ask whether there is a relationship between 
the prices of stocks and bonds. 
The reason why we call this issue a conundrum 
will be readily understood by taking a look at Fig.1 and 2. Let us
comment them briefly (we will come back to them later on). 
\qL
For bonds
it is their yields rather than their prices which is usually
recorded. The yield is the real (as opposed to the coupon rate) interest
rate brought by the bond; it is defined as the ratio of the coupon
rate to the bond's price (for more detail see Appendix A). This means
that over short time intervals of the order of a few months during
which the average coupon rate of a sample of bonds remains approximately
constant, the yield basically represents the inverse of the price. 
Figure 1 shows that for three episodes marked by a stock price dip followed
by a rebound, the yield of US Treasury bonds closely follows the
stock index; in each case the correlation is highly significant
\qfoot{The three correlations in Fig.1
are equal to 0.51, 0.68 and 0.75 
respectively, which for $ n>50 $ are of course highly significant;
subsequently,
for the sake of simplicity, confidence intervals will 
be omitted except in those cases where they matter that
is to say for fairly weak correlations. We computed 
the correlations for the prices themselves; 
as no trend has to be removed
using price changes instead, would present no real advantage
here; 
the main effect would be to increase the noise component.}
. 
In 
other words, prices of stocks and bonds move in opposite directions. 
How can we account for that observation? A first explanation, 
one which may sound
particularly appealing to physicists, is to say: let us assume that
the stock exchange is a closed system (which is not altogether absurd
over short time spans) and for the sake of simplicity let us discard
all other products except stocks and bonds, then the stock exchange 
would be fairly well described by a communicating vessel model in
which the amount of liquid represents the total amount of capital while
the levels of the liquid would represent price levels.
In such a model if the level in vessel  $ S $ goes up,
the level in vessel $ B $ must go down, which is what we observe.
This argument can be fleshed out by remarking that investors usually 
respond to a sharp drop in stock prices by a collective flight to
safety, by which one means that investors are tempted to sell all 
their risky assets and to seek refuge into non risky assets such
as Treasury bills and bonds. Subsequently, when the 
panic abates they will transfer back capital from bonds to stocks.
\qL
But these nice explanations, and especially the first one,
do not hold very long if we bring in
additional evidence. If we focus our attention on the time
interval from 1 June 1999 to 11 October 1999, instead of obtaining
a positive correlation between stock prices and bond yields we 
get instead a negative correlation equal to -0.56. Thus, the 
communication vessel model crumbles down. But we can still hope to
safe the flight to safety model. Indeed, as the above time interval
does not contain any crash-rebound episode, the flight to safety 
mechanism simply does not apply. The main problem when trying to
check this effect is to make sure that the evolution in  yield
indeed reflects the one in price, in other words we ought to make
sure that the average coupon rate of the set of bonds that one
considers remains constant. If one relaxes this constraint, then
the connection between stock prices and bond yields disappears 
completely. This is what happens if one considers a broad time interval,
as illustrated in Fig.2. The thin line gives the correlation
of the Standard and Poor's 500 stock index and the Treasury yields.
The average correlation is -0.25 but this figure has in fact little
meaning since the correlation fluctuates wildly. If we focus
at the period 1969-1980 which was marked by several stock price 
slides, we see that instead of being positive the correlation in fact
is consistently negative.
\qpar
  \begin{figure}[p]
    \centerline{\psfig{width=15cm,figure=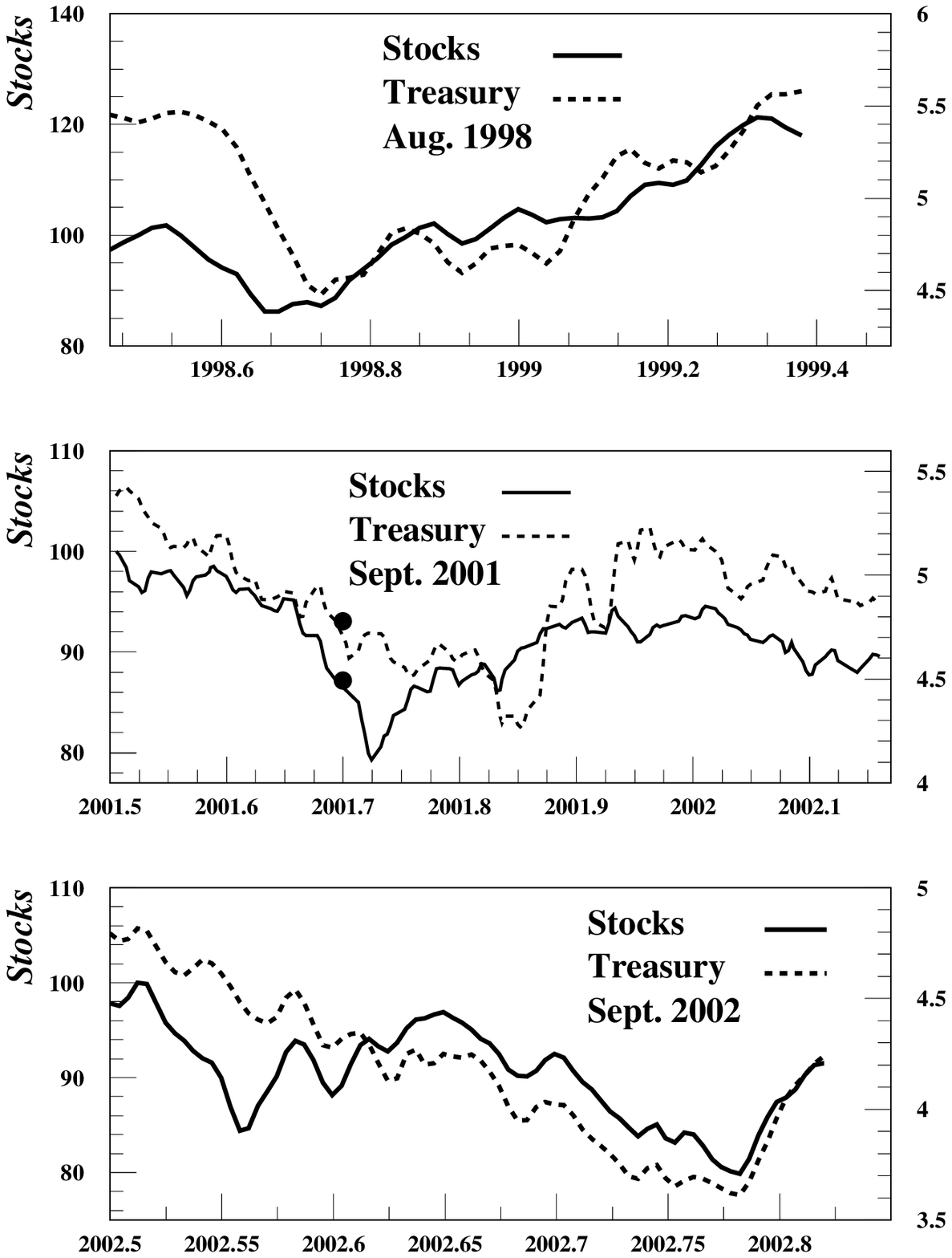}}
    {\bf Fig.1 Comparison of share prices and Treasury yields
during three crashes on the New York stock market.}
{\small The three price falls occurred in the wake of the bull
market of the 1990s. In order to facilitate the comparison the
stock index was normalized to 100 at the beginning of the crashes
(left-hand scale); the right-hand scale represents the yield
of 10-year Treasury bonds in percent. All curves have been 
smoothed through a 3-point moving window average. 
The data are weekly in case 1 and daily in cases 2 and 3.
The correlations
for the three cases are 0.51, 0.68, 0.75 respectively
which are highly significant as confirmed by the following 
confidence intervals (0.27,0.69),(0.63,0.83),(0.44,0.73).}  
{\small \it Source: http://finance.yahoo.com}.
 \end{figure}

Figure 2 also displays the correlation between stock prices and
bond yield spreads. The spread can be defined in several ways
(see Appendix A); one of the most frequently used is the difference
between Baa rated bonds (a medium quality bond) and Treasury bonds.
The more troubled the economic situation, the less one would expect
investors to invest in risky bonds, and therefore the higher the gap
between the yields of risky and non-risky bonds. 
In other words, the spread can
be seen as a measure of economic uncertainty perceived by
investors, as was proposed in an earlier paper (Roehner 2000).
Naturally, one would expect economic uncertainty to increase
dramatically in the wake of major stock market crashes, in other
words one would expect the spread to expand in times of falling
stock prices. 

  \begin{figure}[tb]
    \centerline{\psfig{width=17cm,figure=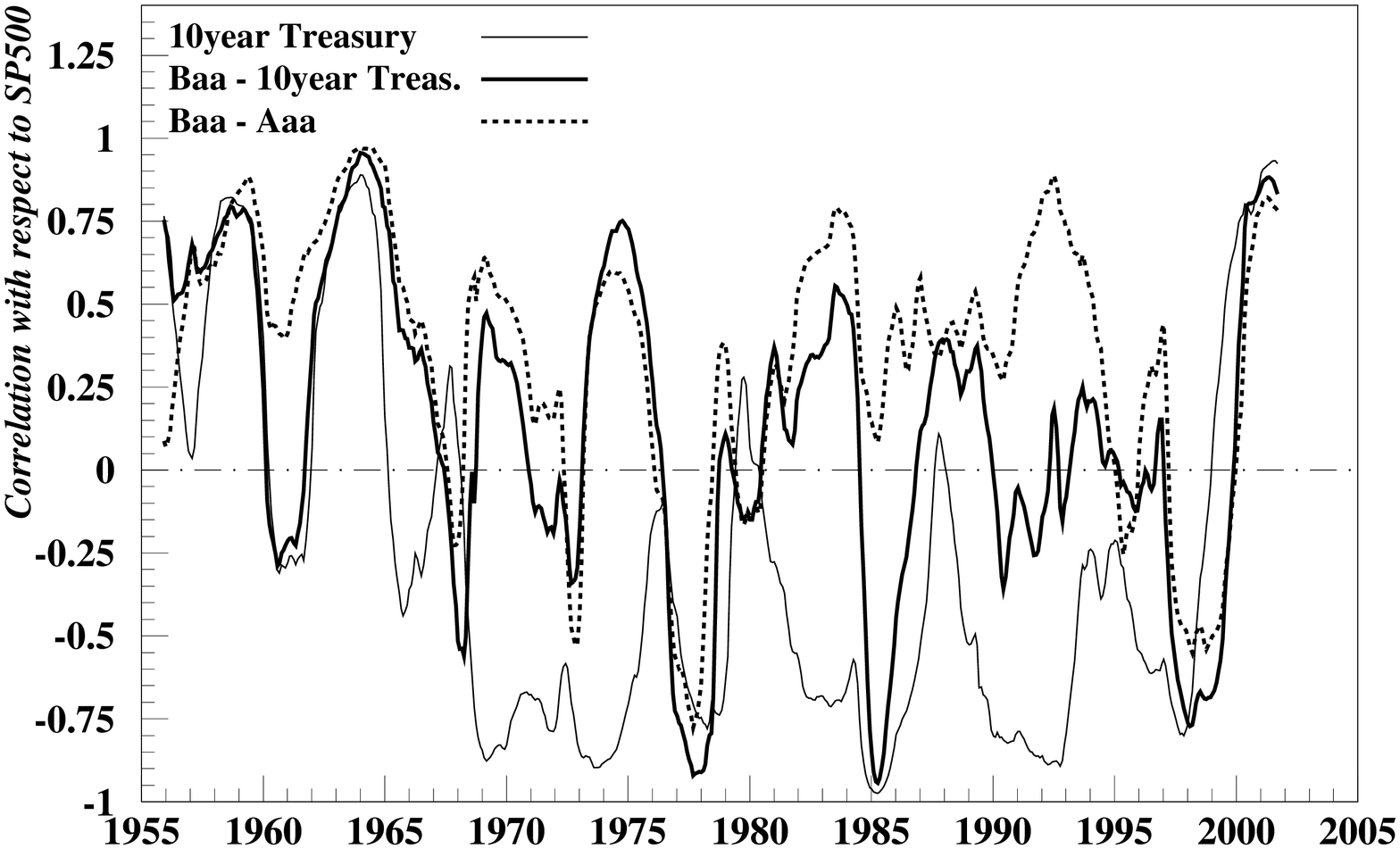}}
    {\bf Fig.2: Correlation between stock prices and bond yields/spreads}.
{\small The correlations were computed using a moving window technique
(width of the window was 41 months). The fact that they fluctuate
wildly shows that there is no stable relationship between stock
prices and bond yields.
Thin line: correlation with
10-year Treasury yields; thick line: opposite of the
correlation of stock prices and the difference
between the yield of corporate bonds of Baa rating and 10-year Treasury
yields. Dashed line: same as thick line except that the 10-year Treasury
yield is replaced by the yield of Aaa bonds, the highest quality
in the range of corporate bonds. Over the whole 1954-2003 time interval
the correlations are -0.25, 0.29 and -0.26 respectively}.
{\small \it Source: http://economagic.com; http://finance.yahoo.com}.
 \end{figure}
A confirmation of this
interpretation can be found in Fig.3: we see 
that the spread widens during stock market crashes and 
narrows during the rebound; over this 7-month interval the correlation
of stock prices and spread is $ -0.85 $.
%
  \begin{figure}[tb]
    \centerline{\psfig{width=17cm,figure=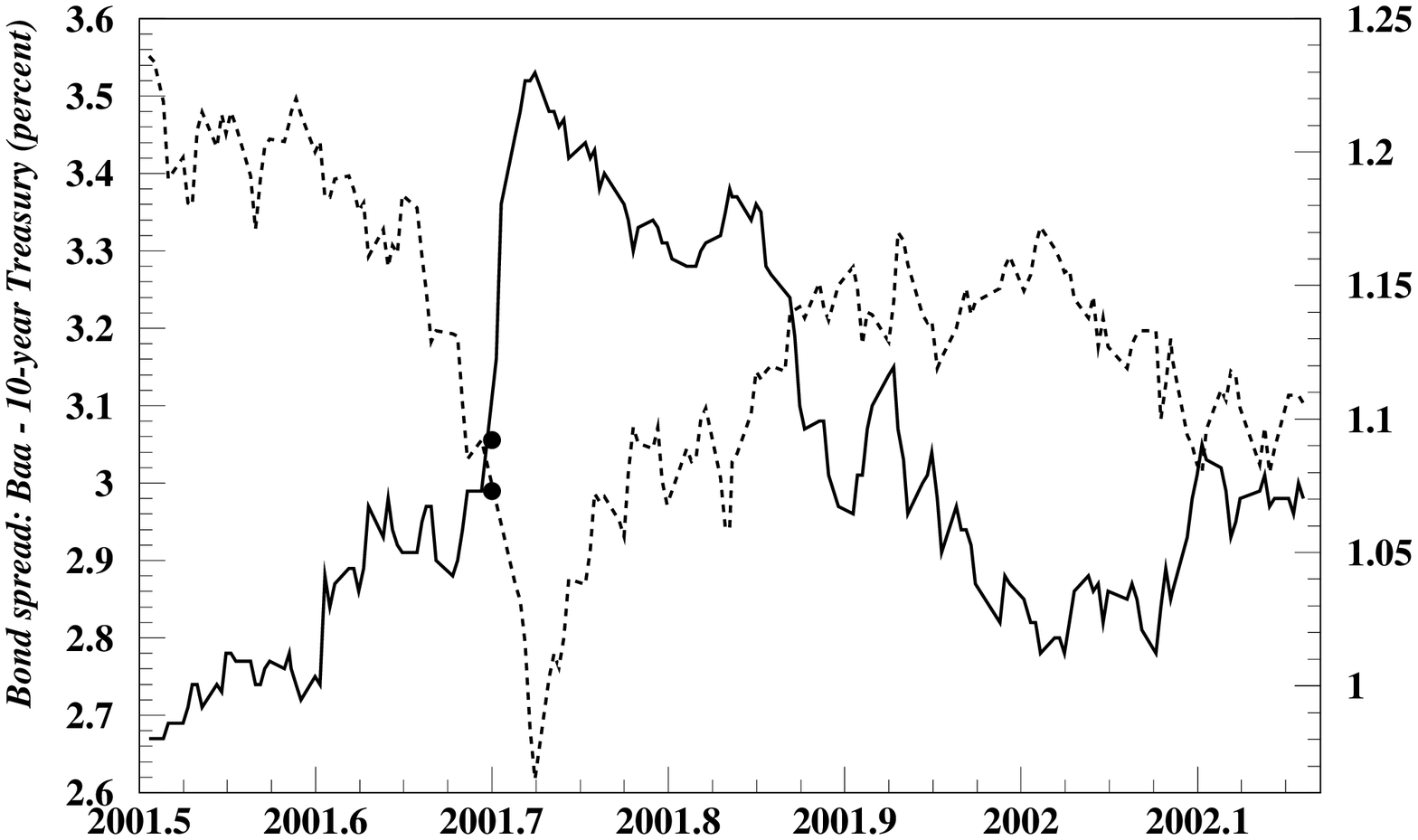}}
    {\bf Fig.3: Bond spread and stock prices}.
{\small Solid line: Difference in the yields of Baa bonds and 10-year
Treasury bonds (left-hand scale); dashed line: Standard\&Poor's 500
index (right-hand scale, divided by 1,000). Daily data; 
the correlation is 0.85. The two dots correspond to September 11}.
{\small \it Source:http://economagic.com; http://finance.yahoo.com}.
 \end{figure}
%
One could expect that 
because it is a {\it difference} of yields the spread would be less 
sensitive to variations in interest rates than the Treasury yield itself.
Yet, Fig.2 shows that it nevertheless
has no stable relationship with stock prices
as shown by the wild fluctuations of the correlation. However the 
spread does a better job in that respect than the Treasury yield.
For instance even a short lived crash such as the one in October 1987
brings about a positive correlation (note that in
order to facilitate the comparison, the graph displays the opposite
of the correlation). However the fact that the overall correlation
for the whole time interval
does not have the right sign (it is equal to 0.29) shows that
it only makes sense to study it if one restrict oneselves to 
selected crash-rebound episodes. 
\qL
This is something economists are reluctant to do. One reason
for their  unwillingness to focus on specific episodes is probably
due to the fact that if one accepts the assumption of a ``rational
homo economicus'' whose way of reasoning would be independent of 
any social {\it Zeitgeist}, it is difficult to imagine that investors
may react differently in an optimistic environment or 
in a time of panic%
\qfoot{A more detailed analysis of this important point can be found
in Roehner 2002 (chapter 3)}%
.
As a result the methodology commonly used in econometrics does not
prove very pertinent for handling problems of 
the kind considered in this paper. 
In standard econometric procedure, time series are treated en bloc
without any attempt to break them up into episodes corresponding to
different mechanisms; once the ``en bloc'' option has been adopted
it makes little difference whether one uses multivariate analysis or any
other statistical tool for the data are spoiled from the start.
A recent paper by Gabe de Bondt (2002) is typical of this procedure.
The author considers (en bloc) the three-year time interval
since the introduction of the euro in
January 1999 and develops a multivariate analysis with no less than
25 variables among which the corporate debt, price earnings ratio,
corporate bond spreads, industrial confidence indicator, etc.
With such a catch-all set of variables one is not in the best
conditions for identifying and isolating a specific mechanism.
In contrast, in a very stimulating paper which is one of the
few exceptions to the above procedure, 
Frederic Mishkin (1991) focuses on a set 
of sharply defined banking panics.
\qpar

The approach that we use in this paper consists in observing the
response of the system to different kind of shocks. In physics,
the perturbations used in order to probe a system can be selected
at will by the experimentalist. Here, we are more in the situation
of an astrophysicist who has to wait for a new supernova to appear
in order to use it as an observational probe. In the next section
we study how the system responded to a series of historical crashes.
Then, in section 3, we take a close look at how it reacted to the
shock of September 11, 2001. In the next section we
study the transition between the crash-rebound regime and
the more ``normal'' regime characterized by the absence of collective
panic reactions. 
Finally, in the last section we
propose some testable predictions. We do not claim that our study
altogether solves the problem, but it provides a series of robust
regularities which should be of usefulness in future attempts to
build mathematical models or simulations. 

\qI{Response to historical stock market crashes}

So far we have considered two variables: the interest rate of
Treasury bonds and the spread of corporate bonds. By observing 
the response of the US market to a sequence of nine historical stock
market crashes, we will be able to determine which one of these variables
has the most robust relationship with stock prices. Fig.4 represents
the correlation of stock prices and interest rates (triangles) or
yield spreads (circles). Obviously it is the spread which has the
most robust relationship. As a matter of fact, such a stable relationship
over a time span of one century and a half is quite remarkable,
especially on account of the major institutional and organizational
transformations that occurred at the New York Stock Exchange during
this time. Although less stable, the pattern displayed by the triangles
is not without interest. 
  \begin{figure}[tb]
    \centerline{\psfig{width=17cm,figure=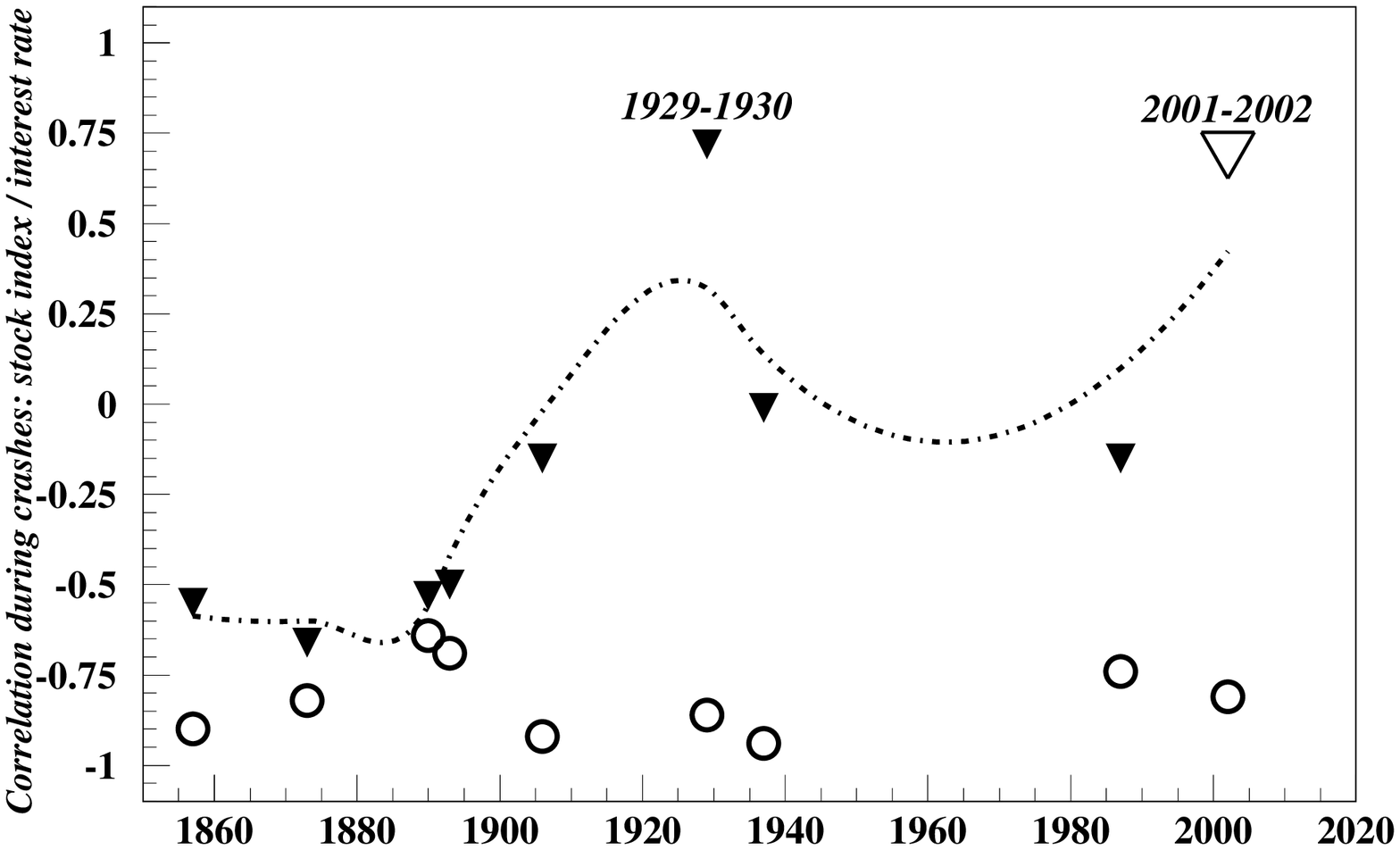}}
    {\bf Fig.4: Correlation between stock prices and interest 
rates/spreads for nine historical episodes of crashes}.
{\small Triangles: correlation between stock prices (SP500) and
the 10-year Treasury yield; the last triangle has been drawn in
white instead of black in order to show that this episode is not
completely over. White circles: correlation between the same stock
index and the Baa-Treasury spread; the graph shows that for the
nine episodes this correlation is more stable than the first}.
{\small \it Source: Roehner (2000)}.
 \end{figure}
First, we see that there is an overall 
upward trend. Back in the nineteenth century, interest rates experienced
a jump during stock market crashes; in these times over-investment
provoked a dearth of capital (usually referred to as a credit crunch)
which naturally lead to higher borrowing prices. Apart from this
trend the cases of 1929-1930 and 2001-2002 stand out as being
characterized by high positive correlations. These crashes were both
preceded by a decade-long period of speculative frenzy and in both cases
interest rates were massively reduced by the Federal Reserve (which
had been created in 1913) in order to check the fall in stock prices
(more details about the case of 1929-1930 can be found in Roehner 2001,
p.186-188). For the sake of illustration Fig.5 compares the evolution
of stock prices, spread and interest rate for the episode that 
is currently under way. This figure extends to 2003 the graphs
displayed in Roehner (2000) for the eight earlier episodes.
%
  \begin{figure}[tb]
    \centerline{\psfig{width=17cm,figure=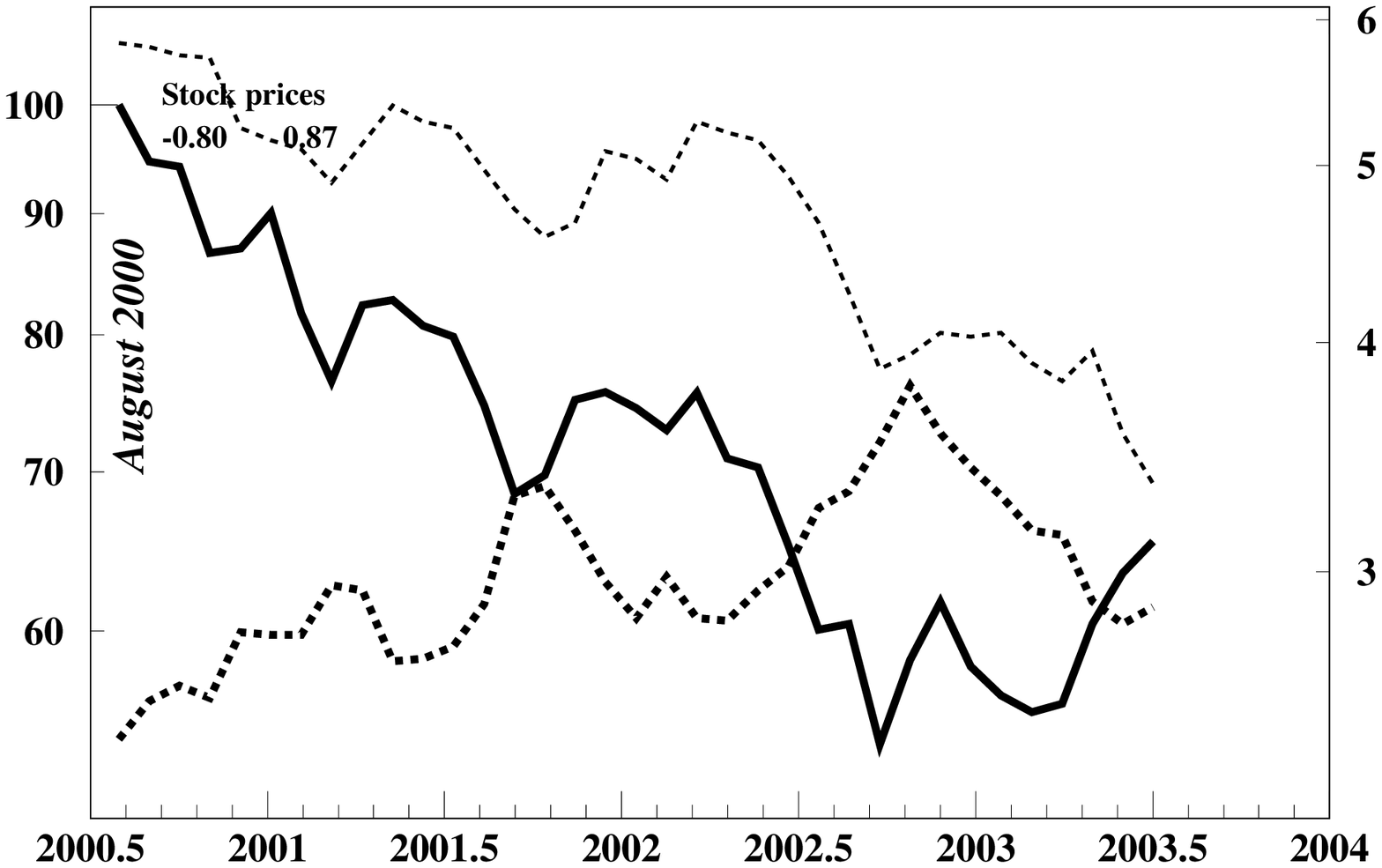}}
    {\bf Fig.5: Stock prices compared with Treasury yield and spread
between start of slide and June 2003}.
{\small Solid line: stock prices (normalized SP500 index),
left-hand side scale. Thin broken line: 10-year Treasury yield;
thick broken line: Baa-Treasury spread, right-hand side scale
in percent for both curves. The correlations are indicated
in the upper left corner. The graphic starts in August 2000 as
indicated by the date printed vertically along the y-axis.
This figure parallels and continues the
graphs given in Roehner (2000) for eight former historical episodes.}
{\small \it Source: http://economagic.com; http://finance.yahoo.com}.
 \end{figure}
%
Table 1a provides regression results for the
nine episodes.



\begin{table}[htb]

 \small 
\centerline{\bf Table 1a\ Relationship between interest rate spread
and stock prices in historical crash episodes}
\centerline{\bf in the United States over the time interval 1857-2003:} 
\centerline{$ \Delta \hbox{Spread}/\hbox{Spread}
=a\left( \Delta \hbox{Stock price}/\hbox{Stock price}\right)+ b $ } 

\vskip 3mm
\hrule
\vskip 0.5mm
\hrule
\vskip 2mm

$$ \matrix{
\tvi 
\hbox{}\hfill & \hbox{Year}  & \hbox{Duration} 
& \hbox{Amplitude} &    a & b & \hbox{Correlation} \cr
\hbox{}\hfill & \hbox{}  & \hbox{} 
& \hbox{of fall } &     &  & \hbox{} \cr
\qtb \hbox{}\hfill & \hbox{}  & \hbox{[month]} 
& \hbox{ } &     &  & \hbox{} \cr
\noalign{\hrule}
\qth 
1 & 1857 & 15 & 0.60 & -0.94 \pm 0.4 & -0.8 \pm 3 & -0.82 \cr
2 & 1873 & 15  &0.80  & -2.00 \pm 1  &\phantom{-} 0.2 \pm 4 & -0.78 \cr
3 & 1890 & 18 & 0.80 & -0.82 \pm 0.4 & -0.8 \pm 1 & -0.74 \cr
4 &1893  &12  & 0.73 & -1.70 \pm 0.7 & -3.7 \pm 3 & -0.84 \cr
5 &1907  &22 & 0.62 & -1.00 \pm 0.6 &\phantom{-} 0.1 \pm 3 & -0.57 \cr
6 &1931  & 44 &0.18  & -0.51 \pm 0.3 &\phantom{-} 3.0 \pm 6 & -0.62 \cr
7 &1938  & 22 &0.55  & -0.75 \pm 0.4 &\phantom{-} 1.5 \pm 4 & -0.78 \cr
8 & 1987 & 10 &0.70  & -0.63 \pm 0.6  & -0.9 \pm 5 & -0.63 \cr
9 & 2002?  &34?  &0.55?  & -0.42 \pm 0.3 & -0.2 \pm 2 & -0.43 \cr
 &  &  &  &  &  &  \cr
\qtb & \hbox{{\bf Average}}\hfill &  &  & \hbox{{\bf -0.98}} \pm 0.2
& \hbox{{\bf -0.20}}\pm 1  & \hbox{{\bf -0.69}} \cr
\noalign{\hrule}
} $$

\vskip 1.5mm
Notes: The year refers to the trough of the stock price index. The
amplitude of the fall is defined as the ratio of the stock
price index at the trough to its level at the peak. 
The cases considered for the period before 1990 are essentially
those identified and enumerated in Mishkin (1991). All data
are monthly.
The variation ratios
$ \Delta \hbox{Spread}/\hbox{Spread} $ and 
$ \Delta \hbox{Stock price}/\hbox{Stock price} $ are defined in percent.
The fact that the average of $ a $ is approximately equal to -1 means
that when the spread decreases by 26 percent (as was the case between
the end of 2002 and mid-2003 one may expect the SP500 to increase
by 26 percent (in fact it increased by 22 percent).
\qL
Sources: Mishkin (1991); http://economagic.com; http://finance.yahoo.com
\vskip 2mm

\hrule
\vskip 0.5mm
\hrule

\normalsize

\end{table}




\begin{table}[htb]

 \small 

\centerline{\bf Table 1b\ Relationship between Treasury yields and 
stock prices in three crashes}
\centerline{\bf in the United States over the time interval 1998-2003:} 
\centerline{$ \Delta \hbox{Stock price}/\hbox{Stock price}
=a\left( \Delta \hbox{Yield}/\hbox{Yield}\right) + b $ } 

\vskip 3mm
\hrule
\vskip 0.5mm
\hrule
\vskip 2mm

$$ \matrix{
\tvi 
\hbox{}\hfill & \hbox{Year}  & \hbox{Duration} 
& \hbox{Amplitude} &    a & b & \hbox{Correlation} \cr
\hbox{}\hfill & \hbox{}  & \hbox{} 
& \hbox{of fall } &     &  & \hbox{} \cr
\qtb \hbox{}\hfill & \hbox{}  & \hbox{[month]} 
& \hbox{ } &     &  & \hbox{} \cr
\noalign{\hrule}
\qth 
1 & 1998 & 11 & 0.84 & 0.19 \pm 0.2 & 0.40 \pm 0.7 & 0.25 \cr
2 & 2001 & 9  &0.80  & 0.22 \pm 0.3  & -0.12 \pm 1 & 0.25 \cr
3 & 2002 & 4 & 0.80 & 0.72 \pm 0.5 & 0.04 \pm 2 & 0.61 \cr
 &  &  &  &  &  &  \cr
\qtb & \hbox{{\bf Average}}\hfill &  &  & \hbox{{\bf 0.38}} \pm 0.2
& \hbox{{\bf 0.11}}\pm 0.7  & \hbox{{\bf 0.37}} \cr
\noalign{\hrule}
} $$

\vskip 1.5mm
Notes: The table focuses on the recurrent crashes (displayed in
Fig.1) which occurred in the wake of the bull market of the 1990s.
For some reason (not well understood yet) these crashes differed
from the other ones (except 1929) listed in Table 1a by the fact that
there was a significant positive correlation between the level of
stock prices and the yield of Treasury bonds. All data are weekly.
The relative variations of stock prices and yields are expressed in
percent.
\qL
Source: http://finance.yahoo.com
\vskip 2mm

\hrule
\vskip 0.5mm
\hrule

\normalsize

\end{table}


 The fact that the average value of the coefficient 
$ a $ is approximately equal to minus one means than a $ x $ percent drop
in the spread would be accompanied by a $ x $ percent increase 
($ x>20 $) in stock
prices and vice versa. This is indeed what has been observed during the
past nine months: between October 2002 and June 2003 (time of writing) 
the spread fell by
26 percent while the SP500 gained 22 percent.
\qpar

Before we leave this section an observation is in order about
the way we measure the interest rate spread. As explained in
Appendix A, it mainly depends upon the data that we
have at our disposal; certainly the best procedure is to
calculate the coefficient of variation of the yields of a 
sufficiently large set of individual bonds. This is the procedure
used by Mishkin (1991) for the pre-1935 episodes for which he
could used Macaulay's data (1938). The other option is to take
the difference between the yield of Baa bonds and of Treasury bonds.
This procedure is used by Mishkin (1991) for the episodes which
occurred after 1950 and is the one that we used throughout this 
paper. However, it is clearly less satisfactory than the 
estimate derived from individual bonds for at least two reasons.
\qbu The decision to grade a bond as Baa is made by rating
agencies. But these agencies can make mistakes. An example
was provided in December 2001 by the bankruptcy of Enron
Corporation. Both Moody's and Standard\&Poor's kept Enron's
bonds at investment grade until just 5 days before it
filed for bankruptcy (Wall Street Journal, 28 March 2002);
a similar episode happened on 19 September 2002 when
the company Electronic Data System announced a drastic
reduction in profits (which provoked a fall of 53 percent
of its share price) 
which took rating agencies completely by surprise; after the 
announcement, they downgraded the company's debt.
\qbu By using the Baa-Treasury difference in fact, we
discard the whole spectrum of low quality bonds between Caa and Baa.
These bonds are issued by companies which are usually smaller
and in a more difficult position; but precisely for these
reasons, they would constitute a more sensible barometer
of the business climate. Using this more sensible indicator
would perhaps permit to explain away some of the outliers.
Unfortunately, so far we were not able to find a comprehensive
data set of individual US corporate bonds.

\qI{The response of investors to greater uncertainty}

The historical evidence reviewed in the previous section highlights
a number of interesting facts. (i) During crash-rebound episodes
stock prices and spreads move in opposite directions; in percentage
terms each change of one variable is mirrored by an identical change
of the other (ii) During the two episodes of 1929 and 2001 the flight
to security of investors was masked by the policy of the central bank.
Indeed such a flight would have resulted in an increase in treasury 
prices and thus a decrease in their yields; but the Fed's policy
of lowering interest rates had of course essentially the same effect;
at this point we have no means which would enable us to disentangle
these two effects%
\qfoot{One might think that by focusing on a time interval during which
there has been no reduction in the fed funds rate one would be able
to observe the other effect alone, but this is not completely right
because the fed funds rate is a short-term rate; the fact that it does
not change does not necessarily mean that medium-
and long-term rates remain unchanged as well.}%
.
\qpar

By observing the day-by-day response of investors to a shock we will
be able to get a better understanding of how they react. 
More specifically,
we will address two questions (i) Are the shift affecting bonds always 
a consequence of what happens on the share market? (ii) What forms take the
flight to security?
\qpar

Most often, in business news, moves affecting the bond market are seen as 
a result of changes in stock prices; one would for instance find 
statements such as ``Today Treasury forfeit previous gains in 
wake of Dow's advance''. Can we really take for granted this direction
of causality? The shock of September 11 gives us an opportunity
to test this assumption. Because the stock market was closed during
a longer time interval than the bond market we can observe how
bonds behave in the absence of stock quotations (Fig.6). 
  \begin{figure}[tb]
    \centerline{\psfig{width=17cm,figure=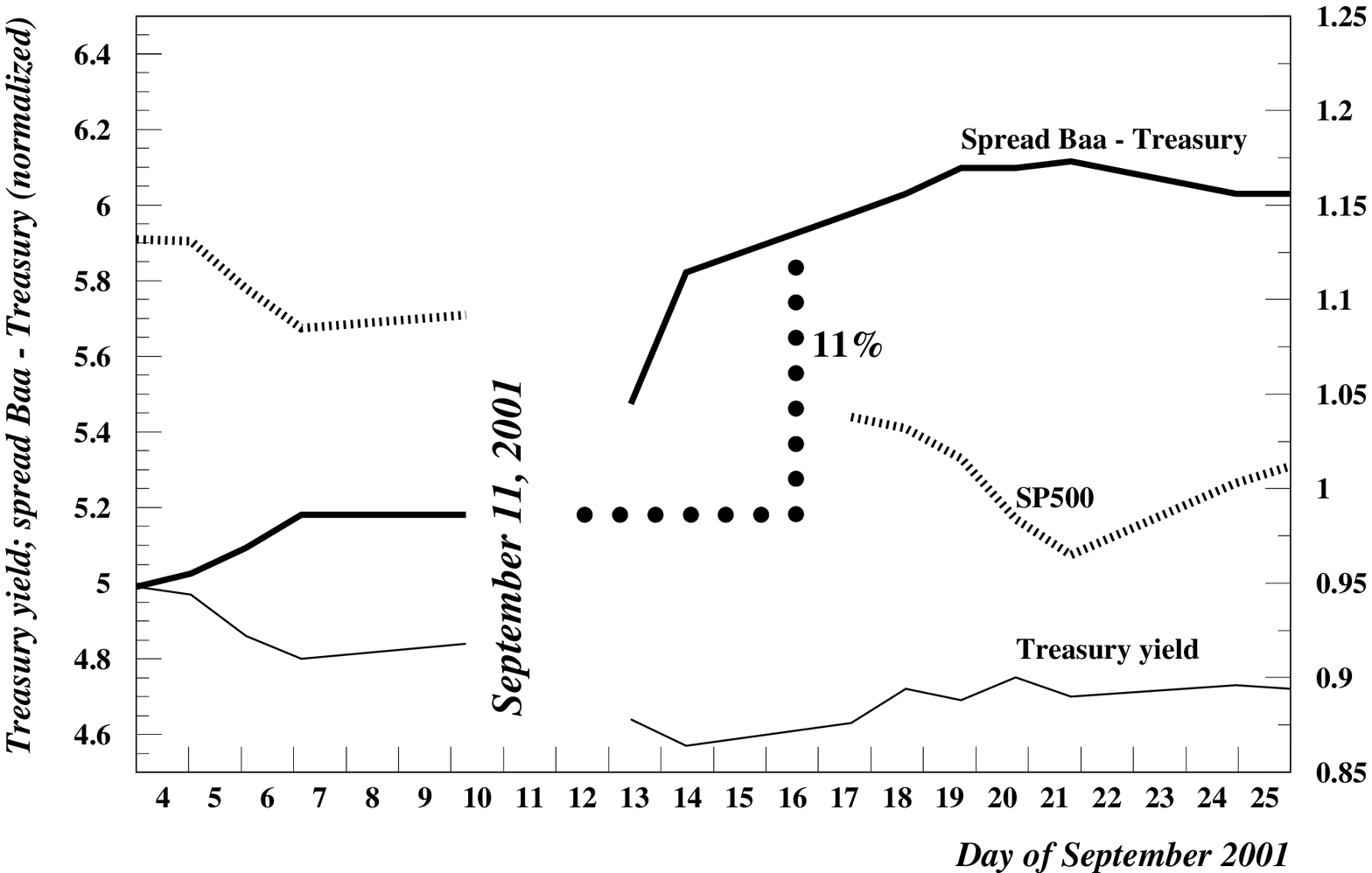}}
    {\bf Fig.6: Response of investors to the shock of September 11, 2001}.
{\small From Sep.10 to its local maximum around Sep.21, the spread 
increased by 18 percent, of which
11 percent occurred before the reopening of stock quotations.}.
{\small \it Source: http://economagic.com; http://finance.yahoo.com}.
 \end{figure}
We observe that:
(i) The 5.6 percent increase in the price of 
10-year Treasury bonds takes place on
September 13-14 that is to say when the stock market was still closed.
(ii) The main part 
(11 percent out of a total shift of 18 percent) of the price fall of
Baa corporate bonds also occurred before the stock market reopened.
In short we see that bond prices can make rapid and
substantial moves without
being pulled by stock prices. 
\qL
Moreover we see that the flight into long-term Treasury
is about two times smaller than the withdrawal from moderately risky
Baa bonds. This conclusion is confirmed by the results given in
Table 2 for bonds quoted in Frankfurt. 


\begin{table}[htb]

 \small 
\centerline{\bf Table 2 \ Response of bonds quoted in Germany to the
shock of September 11, 2001}

\vskip 3mm
\hrule
\vskip 0.5mm
\hrule
\vskip 2mm

$$ \matrix{
\tvi 
\hbox{\bf Baa2 bonds} & & \quad & \hbox{\bf B1 to B3 bonds} \cr
 & & & &  \cr
\hbox{WKN code} & \hbox{Price change} &
& \hbox{WKN code} & \hbox{Price change} \cr
\qtb & \hbox{percent}& & & \hbox{percent}  \cr
\noalign{\hrule}
\qth 610042 & -1.1 & & 352445 & -3.9 \cr
 308960& -1.0 & & 353764 & -3.1 \cr
 352942& -1.3 & &230637  & -9.6 \cr
 610260 & -1.0 & &108565  & -5.3 \cr
 677682& -0.8 & & 614414 & -9.1 \cr
 &  & &  &  \cr
\qtb  \hbox{\bf Average} & \hbox{\bf -1.0} & &  & \hbox{\bf -6.0} \cr
\noalign{\hrule}
} $$

\vskip 1.5mm
Notes: Baa2 bonds are the lower end of the so-called investment-grade
bonds; B1 to B3 bonds belong to the high-risk spectrum; these grades
are just one notch above the Caa grade which designates 
bonds which are close to default. 
It is known that the reaction of bond prices
to changes in interest rates depends upon the times left to maturity.
Here, however, the shock was not primarily a change in interest
rates; nevertheless, we tried to control for a possible effect of
this kind by selecting (as far as available) two sets of bonds with
similar average maturity dates.
\qL
Sources: http://www.finanztreff.de; http://www.bondboard.de (bond finder).
\vskip 2mm

\hrule
\vskip 0.5mm
\hrule

\normalsize

\end{table}


While the withdrawal from
Moody's Baa2 (BBB in Standard and Poor's notation) results in a modest
one percent price decrease, for the B1 to B3 range (Standard and Poor's
B+ to B-) the fall is about 6 percent. 
\qL
If one looks at the change in constant maturity short-term Treasury
bills (3 months to two years) which are not represented on Fig.6 for the
sake of clarity, we see that their prices jump by about 14 percent. 
\qL
To sum up, two major conclusions emerge.
\qbu Bond yields can move quickly and
substantially without such moves having
necessarily to be triggered by shifts in share prices.
\qbu What is usually referred to as a flight to safety in fact
is more a flight away from risk. Investors pull back from what they
perceive as risky assets and transfer the money into secure short-term 
assets such as Treasury bills where it will sit until eventually being 
retransfered to stocks in the rebound phase. 
\qpar

The second point is confirmed by the observation of another shock.
In the afternoon of Tuesday 25 June 2002, the direction of WorldCom,
a major telecommunication company,
announced that due to a multibillion accounting fraud the company
had to fill for bankruptcy. In the following week the SP500 index
fell by about 4 percent; however, for companies in the 
telecommunication sector the fall in share prices 
was much more substantial. For Qwest it was 50 percent;
for Lucent, 33 percent; for Nortel, 23 percent; for IBM it was
only 3 percent; for McDonald's it was less than 2 percent. 
The magnitude of the
drop in fact provides an estimate of how distant the
respective companies are from the telecommunication sector.
What happened to bonds? The price of Baa bonds dropped by 1.8 percent 
while the price of Treasury bonds increased by 0.4 percent. 
\qpar

At this point the reader may wonder whether it is possible to determine
the direction of causality between spread and stock prices by using
statistical methods in the spirit of Granger's causality tests. 
To this aim, we compute the correlation function of the spread and
stock prices by introducing a varying time lag between the two series. 
The rationale of such a procedure is as follows: if the correlation
is found to be maximum for the bond series lagging behind the 
stock prices, one would have good reason to think that the movements
of the stock are the ``cause'' from which changes in  bonds
derive. We performed this test on the daily yield series from
1 July 2001 to 28 February 2002. The results are summarized in table 3.



\begin{table}[htb]

 \small 
\centerline{\bf Table 3 \ Correlation (with time lag $ d $) of spread
and stock prices: Cor[Spread(i),SP500(i+d)]}
\vskip 3mm
\hrule
\vskip 0.5mm
\hrule
\vskip 2mm

$$ \matrix{
\tvi 
\hbox{Time lag} & \hbox{Correlation} & \hbox{Correlation} \cr
\qtb \hbox{[days]} & \hbox{(160-day interval)} & \hbox{(70-day interval)} \cr
\noalign{\hrule}
\qth -6 & -0.800 & -0.881 \cr
-4 & -0.828 & -0.923\cr
-2 & \phantom{^*}-0.852^* & -0.949\cr
\phantom{-}0 & -0.847 &\phantom{^*}-0.954^*\cr
\phantom{-}2 & -0.783 & -0.906\cr
\phantom{-}4 & -0.697 & -0.826\cr
\qtb \phantom{-}6 & -0.610 & -0.734\cr
\noalign{\hrule}
} $$

\vskip 1.5mm
Notes: The series in the second column covers the interval from
1 July 2001 to 28 February 2002; the series in the third column
covers the time interval from 1 July 2001 to 12 October 2001. Both 
time intervals were marked first by a drop in stock prices in the weeks 
preceding and following September 11, and then by a rebound.
The asterisks denote the time lags for which the absolute value
of the correlation is largest.
\qL
\vskip 2mm

\hrule
\vskip 0.5mm
\hrule

\normalsize

\end{table}


We see that the maximum of the correlation occurs for $ d=-2 $ which
corresponds to the spread lagging two days behind stock prices. 
However this test is not completely satisfactory for two reasons
(i) the maximum is fairly soft (ii) the position of the maximum is not
robust with respect to a reduction in the length of the time interval
(column 3 of table 3). In conclusion we can say that the test is
consistent with changes in the two series occurring almost on the same
day. Furthermore one cannot exclude that under a given set of
circumstances the bonds drive the stocks while under different conditions
the stocks drive the bonds. As an image consider the case of a person
who walks his dog; in normal circumstances the dog follows its master,
but if it spots another dog nearby, the situation may well get reversed!

\qI{When the connection between stocks and bonds fades away}
The conundrum evoked in the title of the paper refers to the startling
difference between crash-rebound episodes and long-term behavior. 
The former is marked by a strong connection between stock and
bond prices whereas over the long-term there seems to be no connection
at all. In this section we study more closely the transition between
these two regimes. To this aim we use again the moving window technique
already used in Fig.2 but in a slightly different form.
We start with a time interval marked by a 
strong connection between stock and bond prices and we then progressively
expand this interval by allowing its right-hand boundary to shift toward
24 May 1999.
The result is summarized in Fig.7. 
%
  \begin{figure}[tb]
    \centerline{\psfig{width=17cm,figure=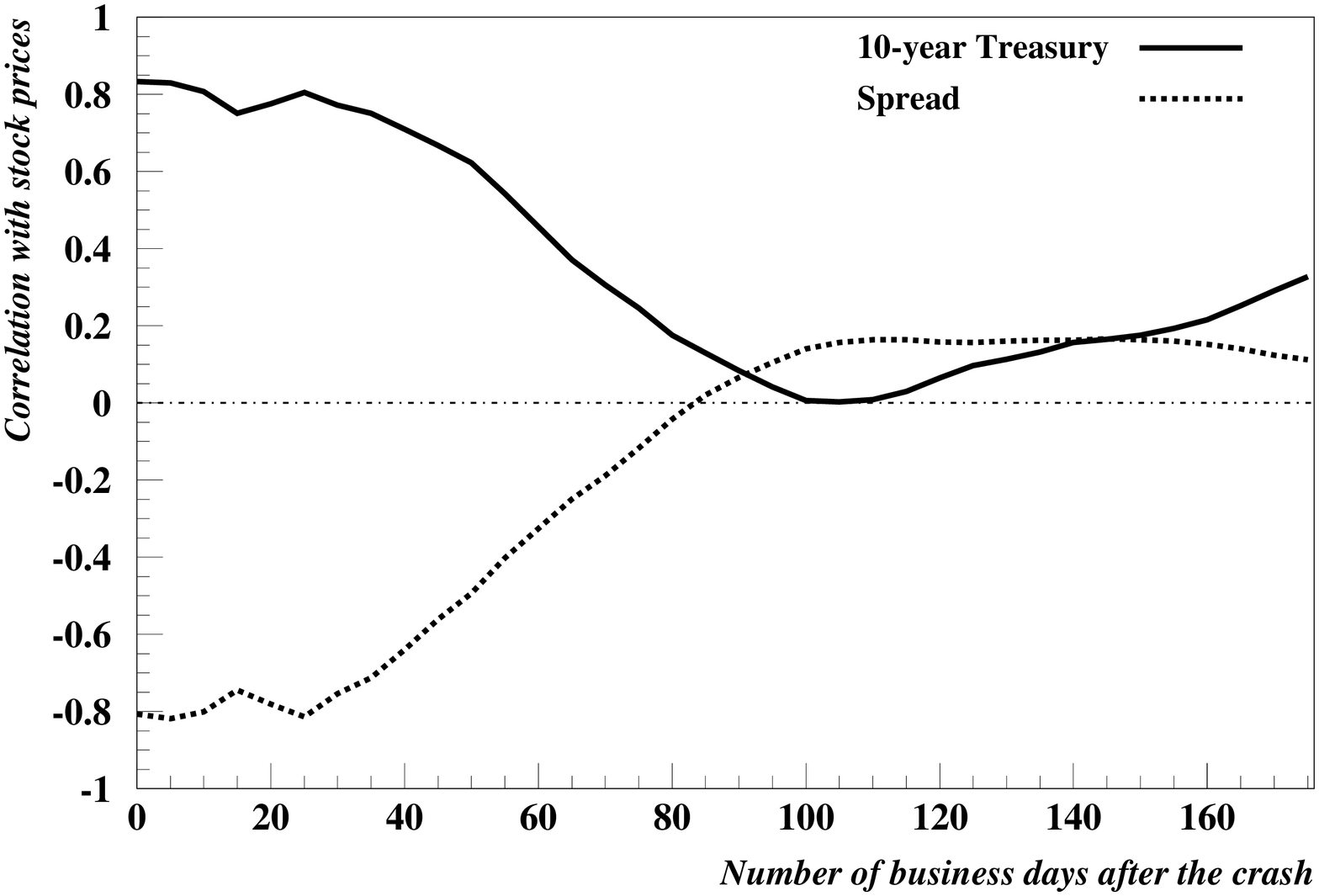}}
    {\bf Fig.7: The correlation between either yield or spread
fades away in the course of time after the crash}.
{\small The decreasing correlation clearly points to a drastic
change in the behavior of investors; it can be interpreted
as a shift away from the collective flight to safety reaction that
prevailed during and shortly after the crash. The graph refers
to the crash of August 1998 (Fig.1a)}.
{\small \it }
 \end{figure}
The time interval is centered
around the crash of August-September 1998; we selected this crash
because in this case the rebound was followed by a long period of
stock price increase (in contrast, after 2000, crashes occurred in
short succession). As the time interval is progressively widened
the two correlations first remain constant, then began to fall 
about 30 business days (i.e. 40 calendar days) after the trough of
the crash; at about 60 days (i.e. 2.6 calendar months)
the correlations are reduced to the point
of being no longer significant. Yet, if we look at Fig.1a it is
difficult to see any difference in the way stock prices progressed. 
The fact that until the end of November 1998 the increase was in fact
a rebound and after that date gave way to the continuation of the
bull market is by no means apparent on the graph of the SP500;
nonetheless there
was a drastic, hidden change in the behavior of investors.  

\qI{Concluding comments}

There are much closer ties between the stock and bond markets than
for instance
between the stock and housing markets if only because there
are securities which provide links between them: for instance preferred
stocks (see Appendix A), despite their name, share many attributes
of bonds, and convertible bonds (see Appendix A) can be transformed into
stocks under pre-determined conditions. Therefore the question of the
connection between stock and bond prices is both a natural and
important one.  We have shown that there is a strong connection 
between stocks and bonds during crash-rebound episodes. Immediately
after the crash, investors sell their risky bonds; the more risky, 
the more they sell them (Table 2); after the rebound they sell some 
of their Treasuries and buy back stocks as 
well and some of the more risky bonds (Fig.1 and 3). 
Subsequently, the collective behavior of investors becomes
less coherent and as other factors (such as for instance changes
in interest rates) take over, the strong connection between stocks and
bonds fades away (Fig.7). 
\qpar

We would like to address three additional points which may be of 
interest in the perspective of future work on this issue. 
\qbu First, why did we restrict this study to US data? Two other possible
candidates would be Europe and Japan. Europe has an important and 
fairly liquid Eurobond market, however European stock markets very much
move in the shadow of Wall Street, a circumstance which is likely
to bias the analysis. The Tokyo stock market is more independent from
New York, but the domestic Japanese bond market lacks liquidity, especially
in the range of medium and low grade bonds. As a matter of fact, it is
because of this lack of liquidity that spread data were not published 
until 1997. Even after this date, daily spread data have fairly large
error bars due to the small number of transactions in low grade bonds;
whereas in 2000 
the turn over (i.e. the ratio of trading volume to capitalization)
of the corporate bond market reached 0.81 in the United States, 
it was equal to 0.16 in Japan (Hattori et al. 2001). Incidentally, the
relative under-development of the Japanese bond market has had 
important consequences during the banking crisis of 1991-2003;
their main assets were stocks and debt of companies, but
after stocks had crashed and the financial situation 
of the companies had deteriorated, there was no junkbond market on
which this bad debt could be sold; as a result it
remained as a millstone around their necks for over a decade.
\qbu Common opinion holds that by lowering short term interest rates
the central bank is able to boost stock prices. However, from the
regularities found in this paper, one could as well made a case for the
opposite to be true. The argument would go as follows. When the 
Federal Open Market Committee lowers the fed funds rate, all existing
bonds (which were issued in an environment of higher rates) become 
more attractive to investors. Consequently their price will climb
while their yield will fall. If this happens in the context of a
crash-rebound episode, we know (Fig.1) that it should be accompanied
by a concomitant fall in stock prices. Thus, we arrive at the conclusion
that cuts in the fed funds rate in fact depress stock prices. 
What makes this reasoning shaky is the fact that it ignores time lags.
As Fig.8 shows, the decisions of the Fed {\it follow} the fluctuations
of short-term interest rates rather than they provoke them. 
Seen in this light, the argument sounds more reasonable. As a matter of
fact, a ``true'' stock market rally should be characterized by a 
transfer of capital from secure assets such as Treasuries to stocks.
In contrast, a rally which occurs amidst increasing Treasury prices
is an anomaly. This was the situation between March and June 2003
(time of writing); in such a case it seems safe to predict that either
Treasury prices will stop increasing or stock will resume their
slide%
\qfoot{The same scenario occurred in August 2002: the SP500 gained
20 percent while the yield of the 10-year Treasury dropped by
13 percent. This rally proved short-lived and was followed by
a 20 percent slide between late August and mid-October 2002. Then in the
last two weeks of October there was a new rally, this time accompanied
by a drop in Treasury prices (a good omen), but also by an increase
in the spread (see Fig.5) which was a less favorable portent.}%
. 

{\bf Predictions} \qL
Ultimately the best way to test models or to check 
the validity of new regularities 
is to propose testable predictions. It is probably no coincidence
that this ``experimental'' procedure has been pioneered by
econophysicists. Marcel Ausloos, Anders Johansen, Didier Sornette
and Nicolas Vandewalle were among the first to test their models
through predictions (see Vandewalle et al. 1998, Johansen et al. 1999,
Sornette 2002).
Naturally, the difficulty of the task strongly depends on whether
one considers short-, medium- or long-term predictions.
For stock price predictions on very short time scales 
(of the order of one minute or less) 
the order book constitute a good guide (see Maslov 2001).
The main difficulty when making predictions over very long
time scales of the order of fifty years or more
(see Sornette 2002, p.373) is that its horizon
may extend beyond the life time of the model.
\qpar

As we already noticed the spread can hardly provide a predictor for 
stock prices. As a matter of fact, short-term prediction of stock
prices is probably impossible anyway for at least two reasons.
Firstly because the market is efficient, a somewhat fuzzy
notion which one can understand in the sense that a multitude of
financial analysts track down any short-term predictor in order 
to exploit it. The second reason is quite different. As is well known,
big companies run massive buyback programs of their own shares. 
As an estimate of their magnitude it is sufficient to recall that
between 1996 and mid-2003 buyback programs represented \$ 1.3 trillion,
that is to say about 13 percent of the capitalization of the NYSE
market at its peak in 2000. Although these programs are 
announced in advance, their timing and the way they are going to be
implemented are not; as a matter of fact,
some are never implemented and are announced merely to reassure
stock holders. Because they are so massive these buybacks
may generate ``spurious'' rallies which of course are impossible to
predict. As another fairly exogenous source of spurious rallies, one
can mention the merger and acquisitions which strongly increase
the demand for stocks but are the result of strategic decisions
made by big companies rather than the consequence of the moves
made by individual investors.
The Toronto stock market offered a spectacular illustration
of this effect over the decade 1991-2000 when merger and acquisitions
were multiplied by a factor 10, before abruptly declining in 2001 and
2002 (Security Industry 2002).
\qfoot{When it comes to short-term predictions about individual stocks
the behavior of the chairman may become a crucial factor. 
For instance, Sanford I. Weill, the chairman of Citigroup, exercised
the right to sell stock options (for more than half a million shares
in each case) on 4 November 2001, 4 November 2002 and 17 June 2003
(http://finance. yahoo.com, section: Citigroup - Insider); Citigroup's
stock price peaked on the first two of these dates, it will be 
interesting to see whether 17 June 2003 will also represent a 
local maximum.
In this
connection it should be remembered that Citigroup owns 10 percent
at least of its stock and has run massive buyback 
programs over the recent years.}
\qpar

However, a buyback rally is unlikely to deeply change the behavior
of investors which means that one can distinguish between genuine
and spurious rally by looking at the changes in bond prices and bond
spreads. If both prices and spreads drop the rally may be genuine.
\qL
This was for instance the case of the rally that followed the crash
of mid-September 2001; unfortunately it did not last very long; basically
it began to fizzle out when stocks had regained their pre-September 11 level.
Looking further into the past, the rally which followed the crash
of late August 1998 was accompanied by a drop in the spread and an
increase in Treasury yields in spite of a cut in the fed funds rate.
Thus, it clearly qualified as genuine. After February 2000 the spread
began to widen very quickly bringing the genuine rally to an end.
However the market stayed on its stride for a while and the downturn
of the S\&P500 occurred only in August 2000.
\qpar

Finally, it must be emphasized that one can of course hardly
expect completely deterministic connections in such a complex
system; it is virtually impossible to control for all the 
variables that we did not consider directly but which may
nevertheless play a role in specific circumstances. 
Fig.2  conveys a feeling of this complexity.
Whereas the stock market is only mildly dependent upon
interest rates, the bond market has a very strong connection
with them, and through these rates it is closely
connected to the ``real economy'' (probably to a greater degree than
the stock market itself). 
Our main objective in this paper
was to scrutinize the behavior of investors during episodes 
marked by a sharp increase in overall uncertainty and to find
regularities in their reactions. Further progress
will become possible if we can find a comprehensive data set
of individual US corporate bonds. 
\qpar

{\bf Acknowledgements \ } We express our gratitude to Dr. Christian
Rauen who introduced us to German electronic data bases covering 
the European bond market; because they are remarkably
user-friendly and provide broad coverage, these data bases
permitted us to perform a number of crucial ``experiments''.
We are indebted to Dr. Yasanobu Katsuki of Mizuho bank for bringing
to our attention crucial characteristics of the Japanese bond market.
\qL
Work at Brookhaven National Laboratory was carried
out under Contract No. DE-AC02-98CH10886, Division
of Material Science, U.S. Department of Energy. B.M.R. thanks
the Theory Institute for Strongly Correlated and Complex Systems
at Brookhaven National Laboratory for financial support during visits 
when part of this work was completed.

 \vfill \eject

 \appendix

 \qI{Appendix A: Some basic facts about bond markets}

This appendix has three purposes. First we recall a few salient facts about
the bond market, then we  highlight its similitudes with the
stock market and finally we describe some recent developments. 
\qpar

First of all one should recall the basic mechanism which governs
the price of bonds. Suppose for instance that a 10-year government
bond was issued by the US Treasury in 1998 with a (fixed) interest
rate of 5 percent which in bond parlance is also called its 
coupon rate. Suppose further that, because there has been a general decline
in interest rates, the coupon rate of this type of bond is
4 percent in 2003; then, at least theoretically, there will be
an extra demand for this older bond with a 5 percent coupon rate
that would cause its price to rise until the ratio of the coupon
rate to the current price (this ratio is called the current yield, 
see more details
below) would be the same 4 percent. In the present case this means
that the price would increase from its initial level of 100 to 
100 times 5/4 that is to say 125.
\qL
As another example, consider a company that issues two different
bonds at the same moment. If their durations are not the same
their coupon rates are also likely to be different. Once they
are on the market, their prices will adjust in the course of time
so that their yields become fairly equal to those of other bonds
of same duration and same quality (see more about quality below) that
are already on the market.
\qpar

Having explained the basic mechanism,
let us now state some of the definitions which are used throughout
this paper. 
\qbu The current yield of a bond is defined as the ratio of 
coupon rate to bond price. By the way, it can be observed that
bond prices are always close to the coupon price which by convention
is taken as equal to 1 or 100 (depending on the convention), except 
for bonds which are close to default. 
The yield to maturity%
\qfoot{Note that in Japan this notion is used with a slightly different
meaning. Whereas in the West, the calculation assumes that the coupon 
payments are reinvested, in Japan interest is not
compounded; this slightly different notion is referred to as the
simple yield (Padua 1998).}
refers to the total revenue gained from the
bond; its computation is more involved because one has to take into
account the present value of all subsequent coupons. Usually, however,
the current yield and the yield to maturity vary in the same direction;
at a qualitative level the two notions can be used without further
distinction.
\qbu Bonds are rated by rating agencies on a scale which comprises
two main classes: the investment grade class which goes from  Aaa to  Baa
in Moody's notation and the high-yield (also called junk bond) class
which goes from Baa to Caa (a Caa grade means that the bond is close
to default)
\qfoot{Here is the whole list of Moody's grades: Aaa, Aa1-Aa2-Aa3, 
A1-A2-A3, Baa1-Baa2-Baa3, Ba1-Ba2-Ba3, B1-B2-B3, Caa.}
. 
The high-yield market first came into existence as a fairly
liquid market in the United States in the early 1980s; in some other
developed countries such as for instance Japan this market is still
fairly narrow and illiquid. 
\qL
Because there is a broad spectrum of grades, a natural question
is to ask how yields depend upon grades. Basically, the lower the 
grade, the higher the yield.
This leads to the notion
of spread. Depending on the data that are available the spread 
can be defined in various ways. If data for a large sample of 
individual bonds are available the spread can be defined
as the (ensemble) standard deviation of the yields. If only average
yield data are available the spread can be defined as the difference
between the yield of Baa bonds and the yield of Treasury bonds. 
Note that although the Aaa grade designates the highest quality
of bonds, Treasury bonds usually are priced higher than Aaa corporate
bonds (for a same coupon rate). This premium is due to several 
factors which make Treasuries more attractive to investors. For instance,
if advantageous to the company, an Aaa bond may be reimbursed before
maturity date, a feature which
introduces additional uncertainty for its owners.
\qpar

When a publicly traded company wants to get financing it has three
options: (i) apply for a bank loan (ii) issue and sell
a bond (iii) issue and sell new stock. 
Once issued and sold to investors, bonds become fairly similar to
shares in the sense that they can be sold and bought and that their
price will fluctuate in the course of time. There are however three 
major differences (i) Year after year, the bond owner is assured to
get a fixed interest rate, the so-called coupon rate. This is why
bonds are called fixed income securities. (ii) At a predetermined
date (the so-called maturity date) the bond owner will be 
reimbursed the face value of the bond. (iii) When the company
fills for bankruptcy, usually holders of common stocks lose 
everything, while bond holders may be able to get at least part of 
their money back. In short, bonds provide predetermined,
regular and secure flows of
income a factor which probably explains that bond prices are much
less volatile than stock prices. When a bond is close to its maturity
its volatility almost drops to zero as the actual price of the bonds
tends toward its face value. 
\qpar
The factor considered as the major determinant of bond prices
is the interest rate. However, as we have seen in this paper, 
bond prices are also affected by many other factors and in particular
by the situation of the stock market. 
\qpar

Interest rates which so to say represent the price of money, are 
crucial (albeit intricate) economic variables which like any other
prices are ultimately determined by supply and demand. This observation
is of little practical interest however for
both the supply and the demand are in fact largely unknown 
(and probably are not even well defined). 
At the short-term end of the spectrum one important
factor is the policy of the central bank. For instance in the United
States, the Federal Open Market Committee meets eight times a year in order
to set the level of the target for the federal funds rate, a rate
which is used for overnight loans to financial institutions. 
Since in this paper we mostly used the benchmark of the 10-year Treasury
bonds it is natural to wonder how this rate is related to the
fed funds rate. Fig. 8 provides a comparison. 
  \begin{figure}[tb]
    \centerline{\psfig{width=17cm,figure=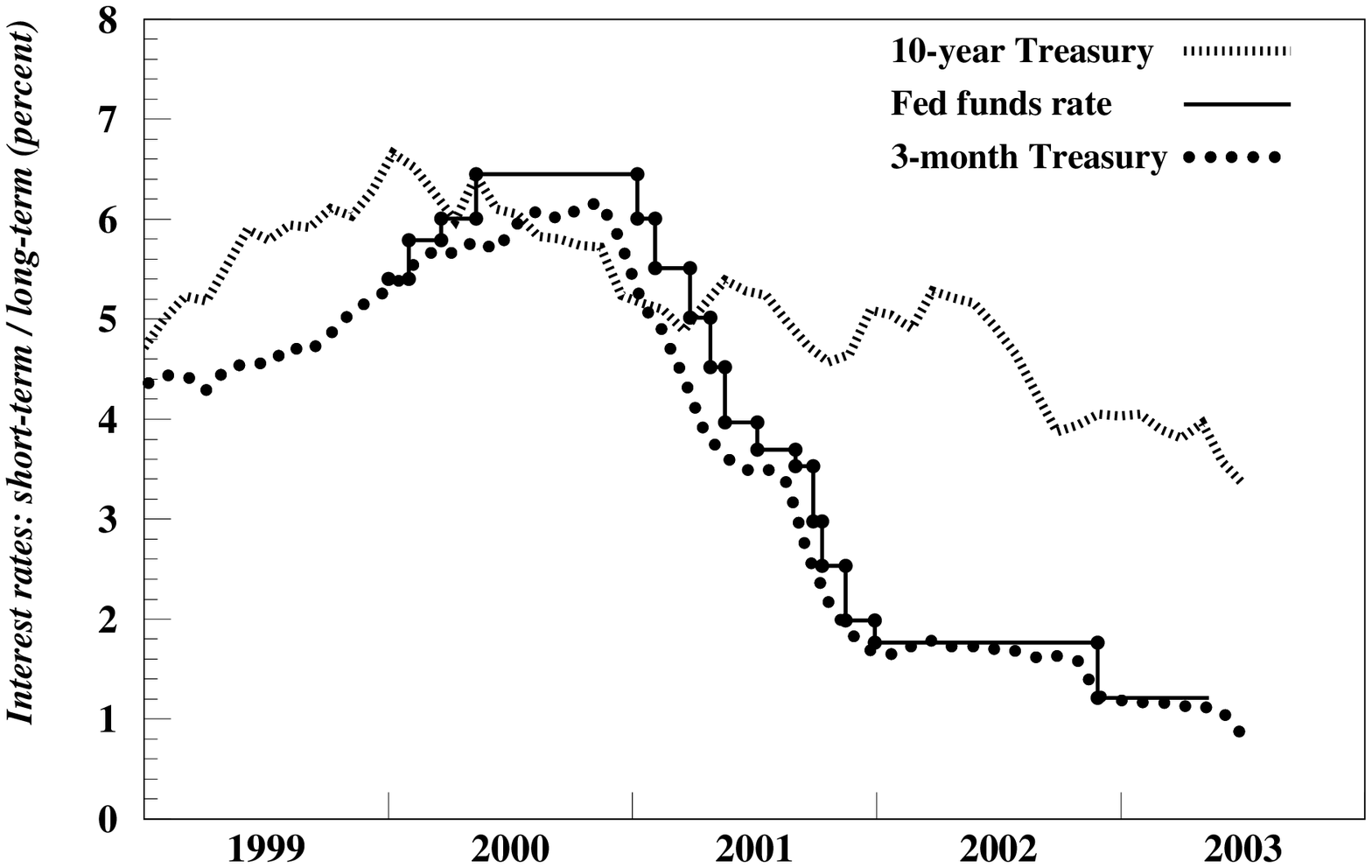}}
    {\bf Fig.8: Short-term and long-term interest rates in the United
States}.
{\small The fed funds rate is a short-term rate that is set periodically
during the meetings of the Federal Open Market Committee. It can be seen
that these decisions in fact follow the changes in the yield of
3-month Treasury bills}.
{\small \it Source: http://www.dallasfed.org; http://economagic.com}.
 \end{figure}
Although there is a close
relationship at the overall level of yearly rates, on shorter time
scales it is not obvious how long term rates derive from short-term 
rates. Fig.8 also emphasizes that the decision of the FOMC follow
the changes in short-term yields brought about by the market rather
than determining them, a feature often overlooked in financial
commentaries. 
\qpar

There is no rigid separation between bonds and stocks. On the one hand,
preferred stocks are similar to bonds in the sense that they carry
no right to vote and in the event of a default their standing is closer
to that of bonds; on the other hand, convertible bonds can be transformed
into stocks under pre-defined conditions. Convertible bonds are in fact
very similar to common stocks in the event of a default. They are
usually issued by companies who would be unable to sell normal bonds;
the convertibility provides a kind of bonus which may attract investors
especially during stock market rallies. In May 2003 convertible
bond issues represented \$ 14 billion, an amount which is of the same
order of magnitude as monthly buybacks. That the linkage 
between stocks and bonds is much stronger than for instance the one
between stocks and real estate is shown by the fact that one cannot
buy a property which would be convertible into stocks (of course
there are real estate investment stocks, the so-called REIT, but
that is purely a stock market affair). 

\vfill \eject

{\large \bf References}

\vskip 5mm

\qparr
Bondt (G. de) 2002: Euro area corporate debt securities market: First
empirical evidence. European Central Bank. Working Papers No 164.

\qparr
Security Industry and Capital Developments, Chartbook 2002: published
by the Investment Dealers Association of Canada. 

\qparr
Hattori (M.), Koyama (K.), Yonetani (T.) 2001: Analysis of credit
spread in Japan's corporate bond market. in: The changing shape of 
fixed income markets: A collection of studies by central bank
economists. Bank of International Settlements (BIS) Papers No 5.

\qparr
Johansen (A.), Sornette (D.) 1999: Financial's anti-bubbles:
log-periodicity in gold and Nikkei collapses. That paper was
first posted as preprint in January 1999 on the Condmat preprint
website (http://xxx.lanl.gov/abs/cond-mat (9901268), then published
in the International Journal of Modern Physics C 10, 563-575.

\qparr
Macaulay (F.R.) 1938: Some theoretical problems suggested by 
the movements of interest rates, bond yields and stock prices 
in the United States since 1856. National Bureau of Economic
Research. New York.

\qparr
Maslov (S.), Mills (M.) 2001: Price fluctuations from the
order book perspective, empirical facts and a simple model.
Physica A, 299,234-246.

\qparr
Mishkin (F.S.) 1991: Asymmetric information and financial crises:
a historical perspective. in R.G. Hubbard ed. Financial markets and
financial crises. A National Bureau of Economic Research Project
Report. University of Chicago Press. Chicago.

\qparr
Padua (A. de) 1998: The international bond market. 
http://www.gwu.edu 

\qparr
Roehner (B.M.) 2000: Identifying the bottom line after a stock 
market crash. International Journal of Modern Physics 11,1,91-100.

\qparr
Roehner (B.M.) 2001: Hidden collective factors in speculative
trading. Springer-Verlag. Berlin. 

\qparr
Security Industry and Capital Developments, Chartbook 2002: published
by the Investment Dealers Association of Canada.

\qparr
Sornette (D.) 2002: Why stock markets crash. Princeton University 
Press. Princeton.

\qparr
Sornette (D.), Zhou (W.-X.) 2002: The US 2000-2002 market descent: how
much longer and deeper? First posted in September 2002
as a preprint on the Condmat website 
(http://xxx.lanl.gov/abs/cond-mat (0209065), then published
in Quantitative Finance 2, 6, 468-481.

\qparr 
Vandewalle (N.), Boveroux (P.), Minguet (A.), Ausloos (M.) 1998:
European Physical Journal B 4,139.

\end{document}